\documentclass[12pt,a4paper,DIV12]{scrartcl}
\usepackage[utf8x]{inputenc}
\usepackage[T1]{fontenc}
\usepackage{lmodern}
\usepackage[british]{babel}
\usepackage{amsmath}
\usepackage{amssymb}
\usepackage{amsfonts}
\usepackage{graphicx}
\usepackage{hyperref}
\usepackage{xcolor}
\usepackage{subfigure}
\usepackage{authblk}
\newcommand*\samethanks[1][\value{footnote}]{\footnotemark[#1]}
\newcommand{\arxiv}[2]{[arXiv:\,\href{http://arxiv.org/abs/#1}
{\texttt{#1}}[\texttt{#2}]]}
\newcommand{\arxivold}[1]{[arXiv:\,\href{http://arxiv.org/abs/#1}
{\texttt{#1}}\,]}
\newcommand{\I}{\ensuremath{\mathrm{i}}}

\newcommand{\tr}{\ensuremath{\mathrm{tr}}}
\newcommand{\mg}{m_{\tilde{g}}}
\newcommand{\api}{\ensuremath{\text{a-}\pi}}
\newcommand{\aetap}{\ensuremath{\text{a-}\eta'}}

\title{%
  {\vspace{-20mm}\normalsize
   \hfill\parbox[b][30mm][t]{35mm}{\textmd{MS-TP-20-17}}}\\[-18mm]
Continuum extrapolation of Ward identities in $\mathbf{\mathcal{N}=1}$ 
supersymmetric SU(3) Yang-Mills theory
}
\author[1,2]{Sajid Ali%
\thanks{\{sajid.ali,munsteg\}@uni-muenster.de}}
\author[3,1]{Georg Bergner\thanks{georg.bergner@uni-jena.de}}
\author[1]{Henning Gerber\thanks{henning.gerber@posteo.de}}
\author[4]{Istvan Montvay\thanks{montvay@mail.desy.de}}
\author[1]{Gernot M\"unster\samethanks[1]}
\author[5]{Stefano Piemonte\thanks{stefano.piemonte@ur.de}}
\author[6]{Philipp Scior\thanks{scior@physik.uni-bielefeld.de}}
\affil[1]{University of M\"unster, Institute for Theoretical Physics, 
Wilhelm-Klemm-Str.~9, D-48149 M\"unster, Germany}
\affil[2]{Government College University Lahore, Department of Physics,
Lahore 54000, Pakistan}
\affil[3]{University of Jena, Institute for Theoretical Physics, 
Max-Wien-Platz 1, D-07743 Jena, Germany}
\affil[4]{Deutsches Elektronen-Synchrotron DESY, Notkestr.~85, D-22607
Hamburg, Germany}
\affil[5]{University of Regensburg, Institute for Theoretical Physics, 
Universit\"atsstr.~31, D-93040 Regensburg, Germany}
\affil[6]{Universit\"at Bielefeld, Fakult\"at f\"ur Physik,
Universit\"atsstr.~25, D-33615 Bielefeld, Germany}
\date{14th May 2020}
\begin{document}
\maketitle

\newpage

\begin{abstract}
\noindent
\textbf{\textsf{Abstract:}}
In $\mathcal{N}=1$ supersymmetric Yang-Mills theory, regularised on a
space-time lattice, in addition to the breaking by the gluino mass term,
supersymmetry is broken explicitly by the lattice regulator.
In addition to the parameter tuning in the theory, the
supersymmetric Ward identities can be used as a tool to investigate lattice
artefacts as well as to check whether supersymmetry can be recovered in the
chiral and continuum limits. In this paper we present the numerical results
of an analysis of the supersymmetric Ward identities for our available 
gauge ensembles at different values of the inverse gauge coupling $\beta$ 
and of the hopping parameter $\kappa$. The results clearly indicate that 
the lattice artefacts vanish in the continuum limit, confirming the 
restoration of supersymmetry.
\end{abstract}

\section{Introduction}

Supersymmetry (SUSY) is an elegant idea which relates fermions and bosons,
whose spin differs by 1/2, through supercharges \cite{Wess-Bagger:1992}. 
SUSY provides dark matter candidates, arising from the lightest supersymmetric 
particles \cite{Jungman:1995df}. In addition to that, supersymmetric
extensions of the Standard Model would resolve the hierarchy problem
\cite{Lykken:2010mc}. $\mathcal{N}=1$ supersymmetric Yang-Mills (SYM) theory,
which is being considered in this article, provides
an extension of the pure gluonic part of the Standard Model \cite{Bergner:2015adz}. 
It describes the strong interactions between gluons and gluinos, the superpartners 
of the gluons. Gluinos are Majorana particles that transform under the adjoint
representation of the gauge group.
The on-shell Lagrangian of $\mathcal{N}=1$ SYM theory, which consists of the gluon
fields $A^a_\mu(x)$ and the gluino fields $\lambda^a(x)$, 
where $a=1, \ldots, N^2_c-1$, can be written in Minkowski space as
\begin{equation}
\label{SYMlagrangian}
\mathcal{L}_{\text{SYM}} = -\frac{1}{4} F^a_{\mu\nu} F^{a,\mu\nu} 
+ \frac{\I}{2} \bar{\lambda}^a \gamma^\mu \left( \mathcal{D}_\mu \lambda \right)^a 
- \frac{\mg}{2} \bar{\lambda}^a \lambda^a,
\end{equation}
where the first term, containing the field strength tensor $F^a_{\mu\nu}$,
is the gauge part, and $\mathcal{D}_\mu$ in the second term is the
covariant derivative in the adjoint representation of the gauge group
SU($N_c$), $N_c$ being the number of colors. The last part of the above
Lagrangian is a gluino mass term which breaks SUSY softly for $\mg \neq 0$,
which means that it does not affect the renormalisation properties of the
theory and that the spectrum of the theory depends on
the gluino mass in a continuous way. 
The physical spectrum of this theory
is expected to consist of bound states of gluons and gluinos, arranged in mass 
degenerate supermultiplets if SUSY is not 
broken~\cite{Veneziano:1982ah,Farrar:1997fn}. 

In order to perform Monte-Carlo simulations of the theory, we discretise the 
Euclidean action and put it onto a four-dimensional
hypercubic lattice. We use the Curci-Veneziano version ~\cite{Curci:1986sm}
of the lattice action $S = S_g + S_f$, where the gauge part
$S_g$ is defined by the usual plaquette action
\begin{equation}
S_g = -\frac{\beta}{N_{c}} \sum_{p} 
\mathrm{Re} \left[ \tr \left( U_{p} \right) \right],
\end{equation}
with the inverse gauge coupling given by $\beta = 2N_c/g^2$, and the
fermionic part
\begin{equation}
S_f =
\frac{1}{2} \sum_{x} \left\{ \bar{\lambda}^{a}_{x} \lambda_{x}^{a} 
- \kappa \sum_{\mu = 1}^{4} 
\left[ \bar{\lambda}^{a}_{x + \hat{\mu}} V_{ab, x \mu} (1 + \gamma_{\mu}) 
\lambda^{b}_{x}
+ \bar{\lambda}^{a}_{x} V^{T}_{ab, x \mu} (1 - \gamma_{\mu})
\lambda^{b}_{x + \hat{\mu}} \right] \right\}
\end{equation}
implements the gluinos as Wilson fermions.
Here the adjoint link variables are defined by
$V_{ab, x \mu} = 2\,\tr\,(U_{x\mu}^\dagger T_a U_{x\mu} T_b)$, where $T_a$
are the generators of the gauge group, 
and the hopping parameter $\kappa$ is related to the bare gluino mass $\mg$
by $\kappa = 1/(2 \mg + 8)$. In order to approach the limit of vanishing
gluino mass, the hopping parameter has to be tuned properly.
In our numerical investigations the fermionic part is additionally
$O(a)$ improved by adding the clover term
$-(c_{sw}/4)\, \bar{\lambda}(x) \sigma_{\mu\nu} F^{\mu\nu} \lambda(x)$
\cite{Musberg:2013foa}.

In our previous investigations we have determined the low-lying mass spectrum 
of the theory with gauge group SU(2) and SU(3)
non-perturbatively from first principles using Monte Carlo
techniques \cite{Bergner:2015adz,Ali:2017iof,Ali:2018dnd,Ali:2019gzj}, and
obtained mass degenerate supermultiplets~\cite{Ali:2019agk}.

\section{SUSY Ward identities}

In classical physics, Noether's theorem provides a relation between
symmetries and conservation laws. In the case of quantum field theories,
symmetries are translated to Ward identities, representing quantum
versions of Noether's theorem. 
In $\mathcal{N}=1$ supersymmetric Yang-Mills theory a gluino mass term 
breaks SUSY softly. The soft breaking effects vanish in the
chiral limit, a limit where theory is characterised by massless gluinos. 
In order to analyse this breaking of supersymmetry and to identify
the chiral limit, we employ the Ward identities for supersymmetry.
Moreover, on the lattice supersymmetry is broken explicitly due to the 
introduction of the discretisation of space-time lattice as a regulator of the theory.
SUSY Ward identities can be used to check whether supersymmetry is
restored in the continuum limit. 

In the Euclidean continuum, on-shell supersymmetry transformations of the gauge and 
gluino fields are given by
\begin{equation}
\delta A_{\mu}^{a} = 
-2\,\I\,\overline{\lambda}^{a} \gamma_{\mu}\, \varepsilon\,,
\quad
\delta \lambda^{a} =
- \sigma_{\mu\nu} F_{\mu\nu}^{a}\, \varepsilon\,,
\end{equation}
where the transformation parameter $\varepsilon$ is an anticommuting Majorana
spinor. From the variation of the action under a supersymmetry transformation 
with a space-time-dependent parameter $\varepsilon(x)$ one derives the SUSY 
Ward identities.
For any suitable gauge invariant local operator $Q(y)$, they read
\begin{equation}
\label{WIs01}
\left\langle \partial^\mu S_\mu(x) Q(y) \right\rangle 
= \mg \left\langle \chi(x) Q(y) \right\rangle 
- \left\langle \frac{\delta Q(y)}{\delta\bar{\epsilon}(x)} \right\rangle ,
\end{equation}
where $S_\mu(x) = (S_{\mu}^{\alpha}(x))$ is the supercurrent of spin 3/2, 
and the term $\mg \left\langle \chi(x)Q(y) \right\rangle$
is due to the gluino mass in the action of the theory.
In the continuum the supercurrent $S_\mu(x)$ and the operator
$\chi(x)$ are given by
\begin{align}
S_\mu(x) &= -\frac{2\,\I}{g} \tr \left[ F^{\nu\rho}(x) \sigma_{\nu\rho} 
\gamma_\mu \lambda(x) \right],\\
\chi(x)  &= +\frac{2\,\I}{g} \tr \left[ F^{\mu\nu}(x) \sigma_{\mu\nu}
\lambda(x) \right].
\end{align}
The last term of Eq.~\eqref{WIs01} is a contact term, which contributes only
if $x=y$, and it can be avoided if $Q(y)$ is not localised at $x$.
Therefore the contact term is ignored in the following discussions.

The four-dimensional space-time lattice breaks SUSY explicitly. 
As a consequence, the lattice versions of the
Ward identities differ from their continuum counter parts by an additional term
$\left\langle X_S(x) Q(y)\right\rangle$. The explicit form of this term 
is known, but need not be displayed here. At tree level this term is proportional 
to the lattice spacing $a$ and vanishes in the limit of zero lattice spacing.
At higher orders in perturbation theory, nevertheless, the contribution of this 
term is finite in the continuum limit due to divergences proportional to 1/$a$ 
that multiply the factor $a$.
This plays a role for the renormalisation of the supercurrent and of the gluino 
mass \cite{Curci:1986sm,Farchioni:2001wx}.
In the renormalisation of SUSY Ward identities, operators of dimensions
$\leq 11/2$ have to be taken into account. They lead to a modification of
the gluino mass, and in addition a current $T_\mu$, mixing with the supercurrent,
appears, corresponding to an operator of dimension $9/2$. 
Consequently, on the lattice the following Ward identities are obtained
\begin{equation}
\label{WIs02}
Z_S \left\langle \nabla_\mu S_\mu(x) Q(y) \right\rangle 
+ Z_T \left\langle \nabla_\mu T_\mu(x) Q(y) \right\rangle
= m_S \left\langle \chi(x) Q(y) \right\rangle + O(a),
\end{equation}
where $Z_S$ and $Z_T$ are renormalisation coefficients. 
The subtracted gluino mass is defined as
$m_S=\mg-\bar{m}$, where $\bar{m}$ is the mass subtraction 
coming from the operators of dimension $7/2$. The mixing current is
defined as
\begin{equation}
T_\mu(x) = \frac{2\,\I}{g} \tr \left[ F_{\mu\nu}(x) \gamma_\nu \lambda(x)
\right].
\end{equation}
Regarding the local insertion operator $Q(y)$, our choice is the spinor
$Q(y)=\chi^{(\mathrm{sp})}(y)$, with
\begin{equation}
\chi^{(\mathrm{sp})}(y) 
= \sum_{i<j} \tr \left[ F_{ij}(y) \sigma_{ij} \lambda(y) \right],
\end{equation}
where the indices $i,j \in \{1,2,3\}$. The reason behind this choice is that
it gives the best signal~\cite{Farchioni:2001wx}.

\section{Numerical analysis of SUSY Ward identities}

We have analysed the SUSY Ward identities numerically, employing the
configurations produced in our project on $\mathcal{N}=1$ supersymmetric 
Yang-Mills theory with gauge group SU(3).
Numerically it is convenient to use integrated Ward identities where
integration or sum is performed over all three spatial coordinates.
The resulting identities will then hold for every time-slice distance $t$. 
In the analysis the data from all time-slice distances in an interval
$t_{min} \leq t \leq t_{max}$ are included.
The lower limit $t_{min}$ is always taken to be larger or equal than 3 in order
to avoid contamination from contact terms. The choice of $t_{min}$ for the
different ensembles of configurations is discussed below.
Since the correlation functions are symmetric or antisymmetric in $t$,
the upper limit $t_{max}$ is chosen to be half of the time extent of the lattice.
Each term in Eq.~\eqref{WIs02} is a 4$\times$4 matrix
in spin-space and can be expanded in the basis of 16 Dirac matrices, 
i.\,e.\ $\left\{\boldsymbol{1},\gamma_5,\gamma_\mu, \gamma_\mu \gamma_5, \I
\sigma_{\mu\nu}\right\}$. It can be shown, with the help of discrete
symmetries, that only the following two contributions are non-zero
\cite{Farchioni:2001wx}:
\begin{equation}
\label{WI2}
\hat{x}_{b,t,1} + A \hat{x}_{b,t,2} = B \hat{x}_{b,t,3},\qquad \text{with}
\quad b=1,2\,,
\end{equation}
where $A=Z_T Z^{-1}_S$, $B=a m_S Z^{-1}_S$, and
\begin{align}
\hat{x}_{1,t,1} &\equiv \sum_{\vec{x}} 
\left\langle \nabla_4 S_4(x) Q(0) \right\rangle,
&\hat{x}_{2,t,1} &\equiv \sum_{\vec{x}}
\left\langle \nabla_4 S_4(x) \gamma_4 Q(0) \right\rangle,\nonumber\\
\hat{x}_{1,t,2} &\equiv \sum_{\vec{x}} 
\left\langle \nabla_4 T_4(x) Q(0) \right\rangle, 
&\hat{x}_{2,t,2} &\equiv \sum_{\vec{x}} 
\left\langle \nabla_4 T_4(x) \gamma_4 Q(0) \right\rangle,\\
\hat{x}_{1,t,3} &\equiv \sum_{\vec{x}} 
\left\langle \chi(x) Q(0) \right\rangle,
&\hat{x}_{2,t,3} &\equiv \sum_{\vec{x}} 
\left\langle \chi(x)\gamma_4 Q(0) \right\rangle. \nonumber
\end{align}
In these equations the Dirac indices of $S_4(x)$, $T_4(x)$, $\chi(x)$ and of the
insertion operator $Q(0)$ are not written, and sums over repeated (hidden) Dirac 
indices are implied. 
Also, $O(a)$ terms that vanish in the continuum limit are not written explicitly
in these equations. Introducing a double index $i=(b,t)$, running over
$2T$ values, where $T$ is the time extent of the lattice, and denoting
$A_1=1, A_2=A, A_3=-B$, Eq.~\eqref{WI2} is written compactly
\begin{equation}
\label{WIs3}
\sum_{\alpha = 1}^3 A_\alpha \hat{x}_{i\alpha} = 0\,.
\end{equation}
In these equations the $\hat{x}_{i\alpha} = \langle x_{i\alpha} \rangle$
are the expectation values of random variables $x_{i\alpha}$, which themselves
are considered to be the results of a finite Markov chain.
We compute the estimators $x_{i\alpha}$ for the correlation functions 
$\hat{x}_{i\alpha}$ numerically using high performance facilities.
The Eqs.~\eqref{WIs3}, including all time-slice distances $t$ from
$t_{min}$ to $t_{max}$, are solved simultaneously for $A_\alpha$ 
by means of minimal chi-squared methods. Two methods, namely the so-called 
Local Method and Global Method, have been used in the past by our 
collaboration \cite{Bergner:2015adz,Farchioni:2001wx}. These methods, however,
do not take properly into account correlations between the different quantities 
appearing in Eq.~\eqref{WIs3}.
For this purpose we have developed a new method based on a generalised least
squares fit, the so-called GLS Method \cite{Ali:2019thesis}, based on the maximum 
likelihood. 
For fixed $A_\alpha$ ($\alpha=1,2,3$) and given numerical data $x_{i\alpha}$, 
the probability distribution $P \sim \exp(-L)$ of the quantities $\hat{x}_{i\alpha}$,
subject to the constraints~\eqref{WIs3}, has its maximum at a point 
where $L = L_{min}$, with
\begin{equation}
L_{min} = \frac{1}{2}\sum_{i,\alpha,j,\beta} (A_\alpha x_{i\alpha})
(D^{-1})_{ij} (A_\beta x_{j\beta})\,,
\end{equation}
where
\begin{equation}
D_{ij} = \sum_{\alpha, \beta} A_\alpha A_\beta (\langle x_{i\alpha} x_{j\beta}
\rangle - \langle x_{i\alpha} \rangle \langle x_{j\beta} \rangle).
\end{equation}
Next, the desired coefficients $A_\alpha$ have to be found such that $L_{min}$
as a function of $A_2$ and $A_3$ is minimised. This cannot be solved analytically,
and we find $A_\alpha$ numerically such that the global minimum of 
$L_{min}(A_2, A_3)$ is reached; for details see Ref.~\cite{Ali:2018fbq}. 
In particular, owing to $A_3= - a m_S Z^{-1}_S$ this provides us with the subtracted 
gluino mass $m_S$ up to the renormalisation factor.
To estimate the statistical uncertainties we employ the standard Jackknife procedure.

\subsection{Discretisation effects}

\begin{figure}[hbt!]
\centering
\includegraphics[width=0.49\textwidth]{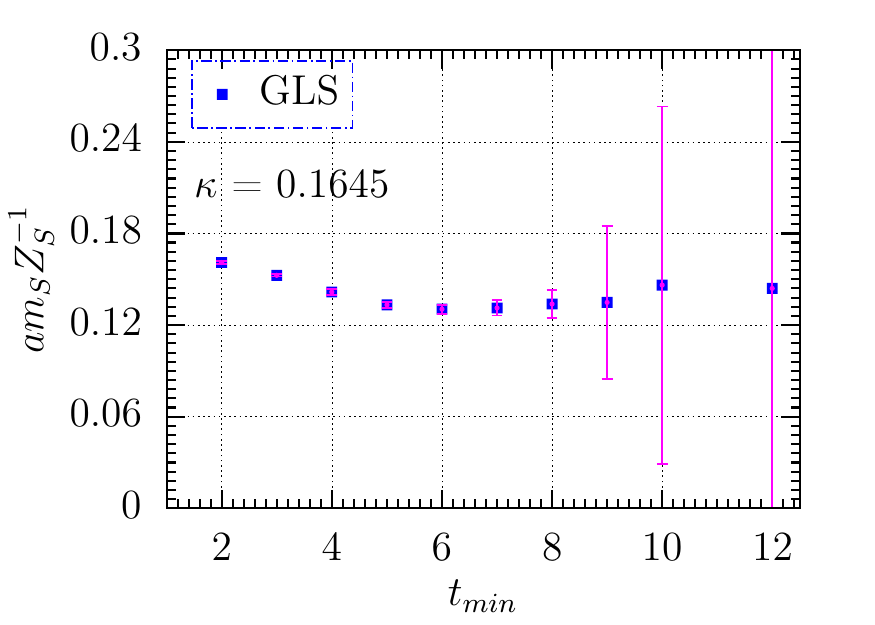}
\includegraphics[width=0.49\textwidth]{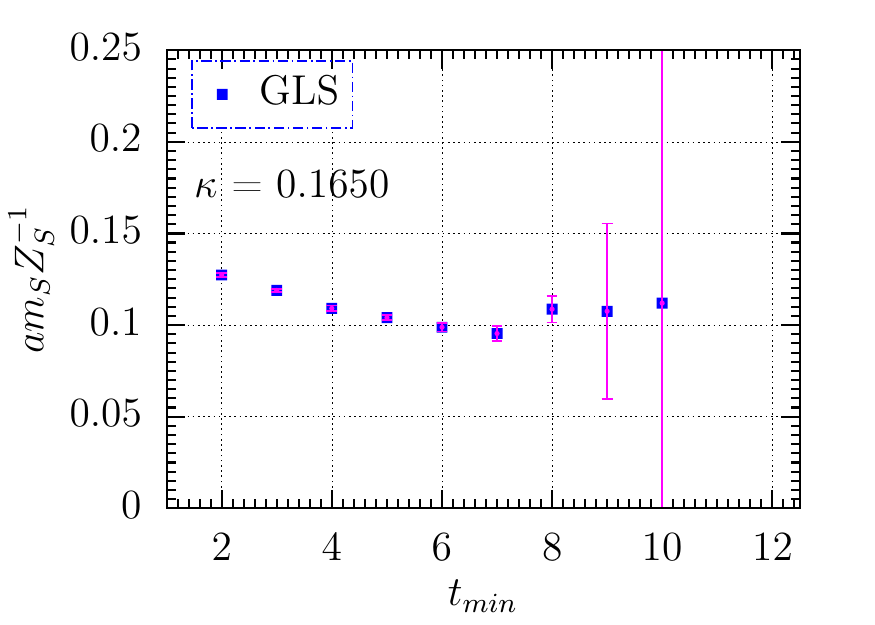}
\includegraphics[width=0.49\textwidth]{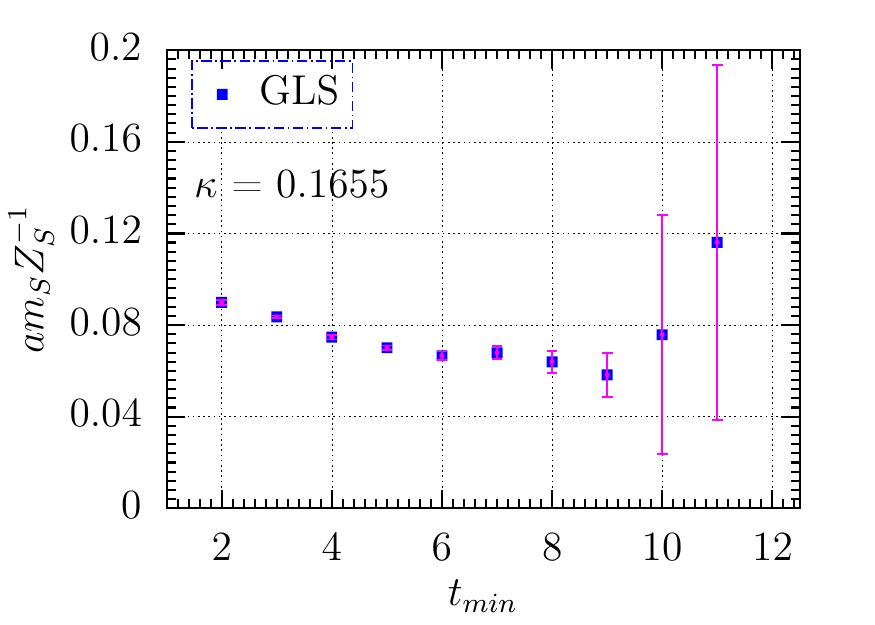}
\includegraphics[width=0.49\textwidth]{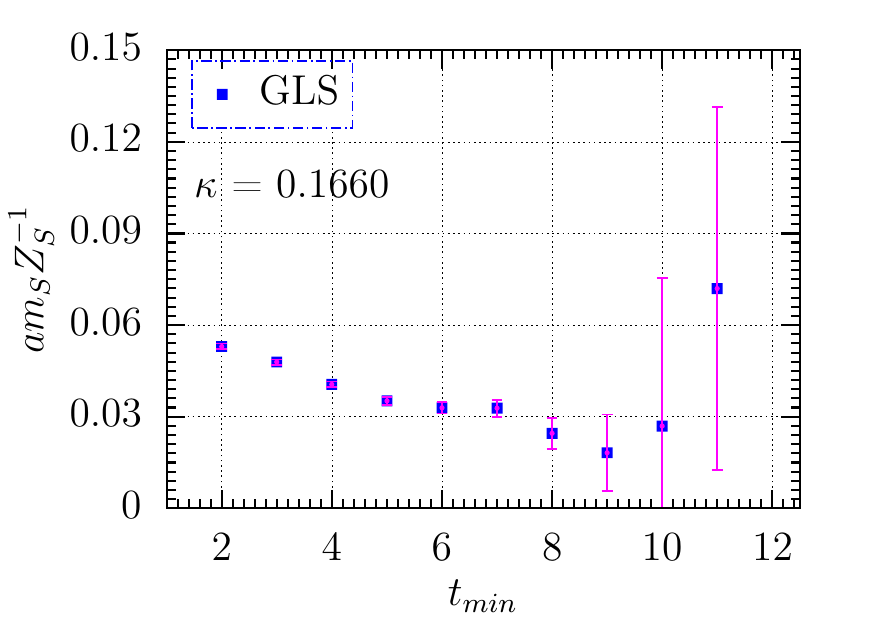}
\caption{The subtracted gluino mass $am_SZ^{-1}_S$ as a function of $t_{min}$ 
calculated with the GLS Method at $\beta=5.6$. At small values of $t_{min}$ 
the subtracted gluino mass is affected by contact terms and by $O(a)$ terms.
Data from $t_{min} = 2$ and $t_{min} = 3$ are shown, but do not enter our 
final analysis.}
\label{FigMgGLSTmin}
\end{figure}

All terms in the Ward identity~\eqref{WIs02}, including the $O(a)$ term
$\left\langle X_S(x) Q(y)\right\rangle$, are correlation functions of gauge 
invariant operators. In the corresponding Eqs.~\eqref{WI2} they are correlation
functions of operators localised on time slices or pairs of adjacent time
slices at distance $t$. As for any gauge invariant correlation function of 
this type, they decay exponentially in $t$, with a decay rate given by the 
mass gap of the theory. For very small $t$ the contributions of higher masses 
will affect the impact of the $O(a)$ term on the Ward identities. 
Therefore we expect that the value of the obtained gluino mass will depend on 
the minimal time slice distance $t_{min}$. This effect should become negligible 
at sufficiently large $t_{min}$. 
On the other hand, if $t_{min}$ is chosen too large, noise in the data
will dominate.
The behaviour that can be observed in Fig.~\ref{FigMgGLSTmin} is compatible
with these expectations.
An adequate choice of $t_{min}$ is therefore important for the quality of the
results. We cope with this in two ways.

In order to avoid perturbing effects at too small $t_{min}$ and a poor
signal-to-noise ratio at too large $t_{min}$, 
for each hopping parameter and inverse gauge coupling,
the value of $t_{min}$ is selected by finding an optimal starting 
point where a plateau in the subtracted gluino mass begins.
The results are presented in Tab.~\ref{TminFormLargeT}.
\begin{table}[hbt!]
\begin{center}
\begin{tabular}{||c|c||c|c||c|c||c|c||c|c||c||}
\hline
\hline
\multicolumn{2}{||c||}{$\beta=5.4$} &\multicolumn{2}{c||}{$\beta=5.4$} &
\multicolumn{2}{c||}{$\beta=5.45$} & \multicolumn{2}{c||}{$\beta=5.5$} &
\multicolumn{2}{c||}{$\beta=5.6$} \\
\hline
\hline
\multicolumn{2}{||c||}{\!\!$V=12^3\times24$\!\!} &\multicolumn{2}{c||}
{\!\!$V=16^3\times32$\!\!} & \multicolumn{2}{c||}{\!\!$V=16^3\times32$\!\!}&
\multicolumn{2}{c||}{\!\!$V=16^3\times32$\!\!} & \multicolumn{2}{c||}{\!\!
$V=24^3\times48$\!\!} \\
\hline
\hline
$\kappa$ & \!\!$t_{min}$\!\! & $\kappa$ & \!\!$t_{min}$\!\! & $\kappa$ &
\!\!$t_{min}$ \!\!& $\kappa$ &\!\! $t_{min}$ \!\!& $\kappa$ & \!\!$t_{min}$\!\!\\
\hline
\hline	
0.1695	& 4 & 0.1692 & 4 & 0.1685 & 5 & 0.1667 & 5 & 0.1645 & 7 \\
0.1700	& 4 & 0.1695 & 4 & 0.1687 & 5 & 0.1673 & 5 & 0.1650 & 7 \\
0.1703	& 4 & 0.1697 & 4 & 0.1690 & 5 & 0.1678 & 5 & 0.1655 & 6 \\
0.1705  & 4 & 0.1700 & 4 & 0.1692 & 5 & 0.1680 & 5 & 0.1660 & 7 \\
-	& - & 0.1703 & 4 & 0.1693 & 4 & 0.1683 & 5 & -	    & - \\
-	& - & 0.1705 & 4 & -	  & - & -      & - & -	    & - \\
\hline
\hline
\end{tabular}
\caption{The values of $t_{min}$ for all available gauge ensembles, 
chosen such that a plateau is formed.}
\label{TminFormLargeT}
\end{center}
\end{table}

In the second approach, we consider that our
simulations of the theory are done at different values of the lattice 
spacing $a$, which leads to different $O(a)$ terms in the Ward identities.
A fixed value of $t_{min}$ in lattice units would mean a lower limit on the
time-slice distances in physical units, that is on the cutoff-scale and 
shrinks to zero in the continuum limit. Instead it would be more appropriate to 
consider $t_{min}$ at constant physical distance for all gauge ensembles.
This is done in the following way.

At the coarsest lattice spacing, at inverse gauge coupling $\beta_0$, 
the value of $t_{min}$ is selected according
to the plateau criterion explained above.
For finer lattice spacings at inverse gauge couplings $\beta_i$
the corresponding $t_{min}$ are then obtained
by scaling with a physical scale.
In order to determine the physical scale we use the
mass $m_{g\tilde{g}}$ of the gluino-glue particle and the Wilson flow 
parameter $w_0$. 
Correspondingly, $t_{min}$ is scaled according to
\begin{align}
t_{min,{\beta_{i}}} &= t_{min,\beta_{0}} 
\frac{m_{g\tilde{g},\beta_{0}}}{m_{g\tilde{g},\beta_i}}\,,\\
\text{or} \quad
t_{min,{\beta_{i}}} &= t_{min,\beta_{0}} 
\frac{w_{0,\beta_{i}}}{w_{0,\beta_0}}\,,
\end{align}
where $\beta_0=5.4$, $\beta_1=5.45$, $\beta_2=5.5$, and $\beta_3=5.6$. 
The resulting $t_{min}$ is rounded to the nearest integer value.
The values obtained by this method are collected in Tab.~\ref{TminfromGGandW0}.
In most points they are equal or almost equal to those in Tab.~\ref{TminFormLargeT}.
\begin{table}[hbt!]
\begin{center}
\begin{tabular}{|c|c|c|}
\hline
$\beta$ & $t_{min}$ from $m_{g\tilde{g}}$ & $t_{min}$ from $w_0$  \\
\hline
5.4	  & 	4	  & 4 \\
5.45	&	5	  & 5 \\
5.5	  &	5	  & 6 \\
5.6	  &	7	  & 7 \\
\hline
\end{tabular}
\caption{The values of $t_{min}$ at fixed physical temporal distance from
scaling with the gluino-glue mass $m_{g\tilde{g}}$ and with the Wilson 
flow parameter $w_0$.}
\label{TminfromGGandW0}
\end{center}
\end{table}
%

\subsection{Adjoint pion and remnant gluino mass}

The chiral limit is defined by the vanishing of the subtracted gluino mass.
Its measured values
can therefore be employed for the tuning of the hopping parameter $\kappa$
to the chiral limit. On the other hand, we can also use the vanishing of the
adjoint pion mass $m_{\api}$ for the tuning~\cite{Demmouche:2010sf}. 
The adjoint pion $\api$ is an unphysical particle in the SYM theory, that can
be defined in partially quenched chiral perturbation theory \cite{Munster:2014cja}.
In the numerical simulations its correlation function can be computed
as the connected piece of the correlation function of the $\aetap$ particle.
Similar to the Gell-Mann-Oakes-Renner relation of QCD
\cite{Veneziano:1982ah}, in the continuum limit there is a linear relation 
between the adjoint pion mass squared and the gluino mass: $m^2_{\api} \propto \mg$.

\begin{figure}[hbt!]
\centering
\subfigure[The subtracted gluino mass $am_SZ^{-1}_S$ and the adjoint pion mass 
squared $(a m_{\api})^2$ as a
function of $1/(2\kappa)$, and the corresponding extrapolations towards
the chiral limit ($\kappa_c$).]
{\label{chiral}
\includegraphics[width=0.46\textwidth]{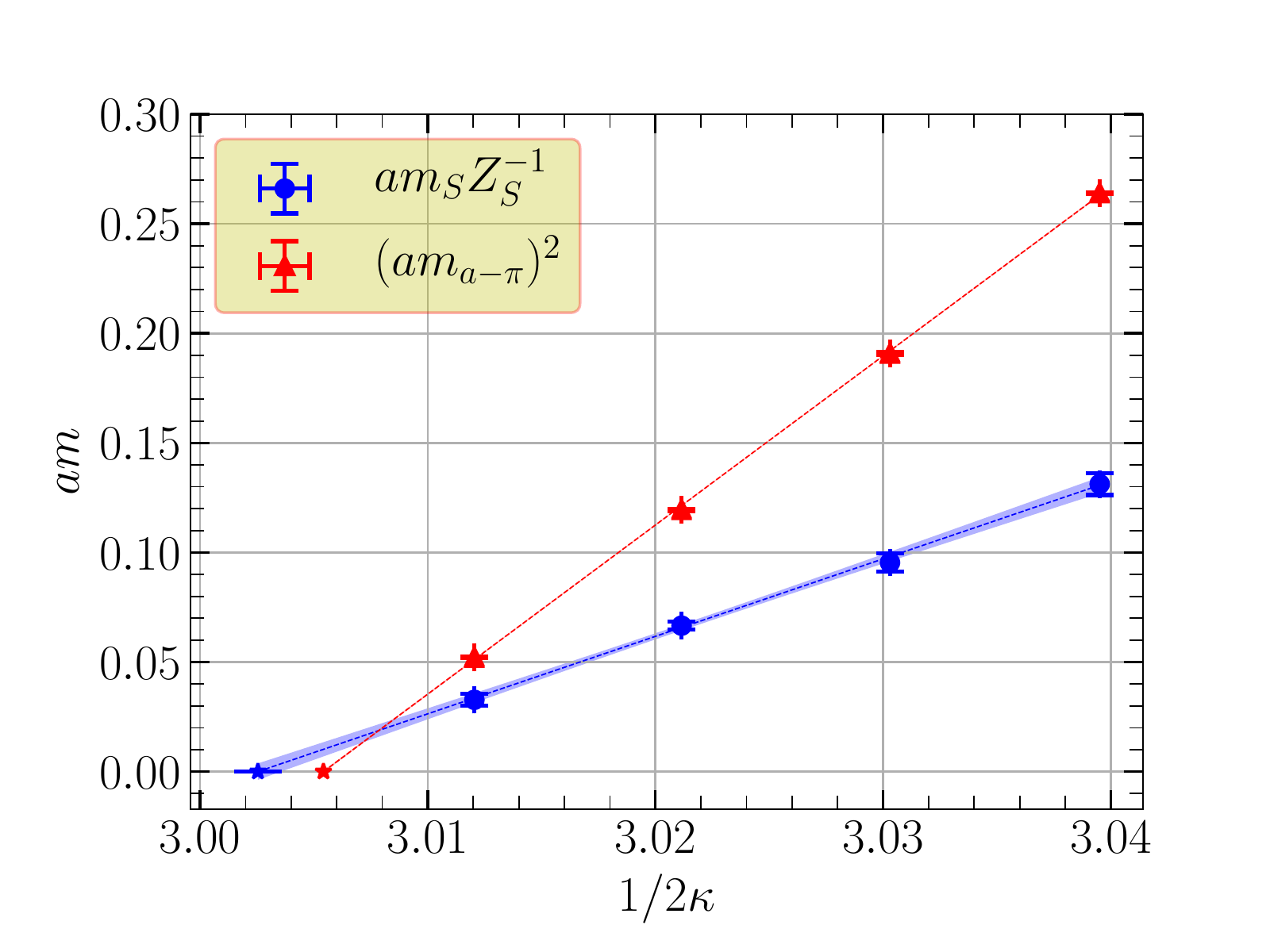}}
\qquad
\subfigure[The subtracted gluino mass $am_SZ^{-1}_S$ as a function of the adjoint 
pion mass squared $(a m_{\api})^2$
in order to obtain the remnant gluino mass $\Delta(am_SZ^{-1}_S)$.]
{\label{subtracted}
\includegraphics[width=0.46\textwidth]{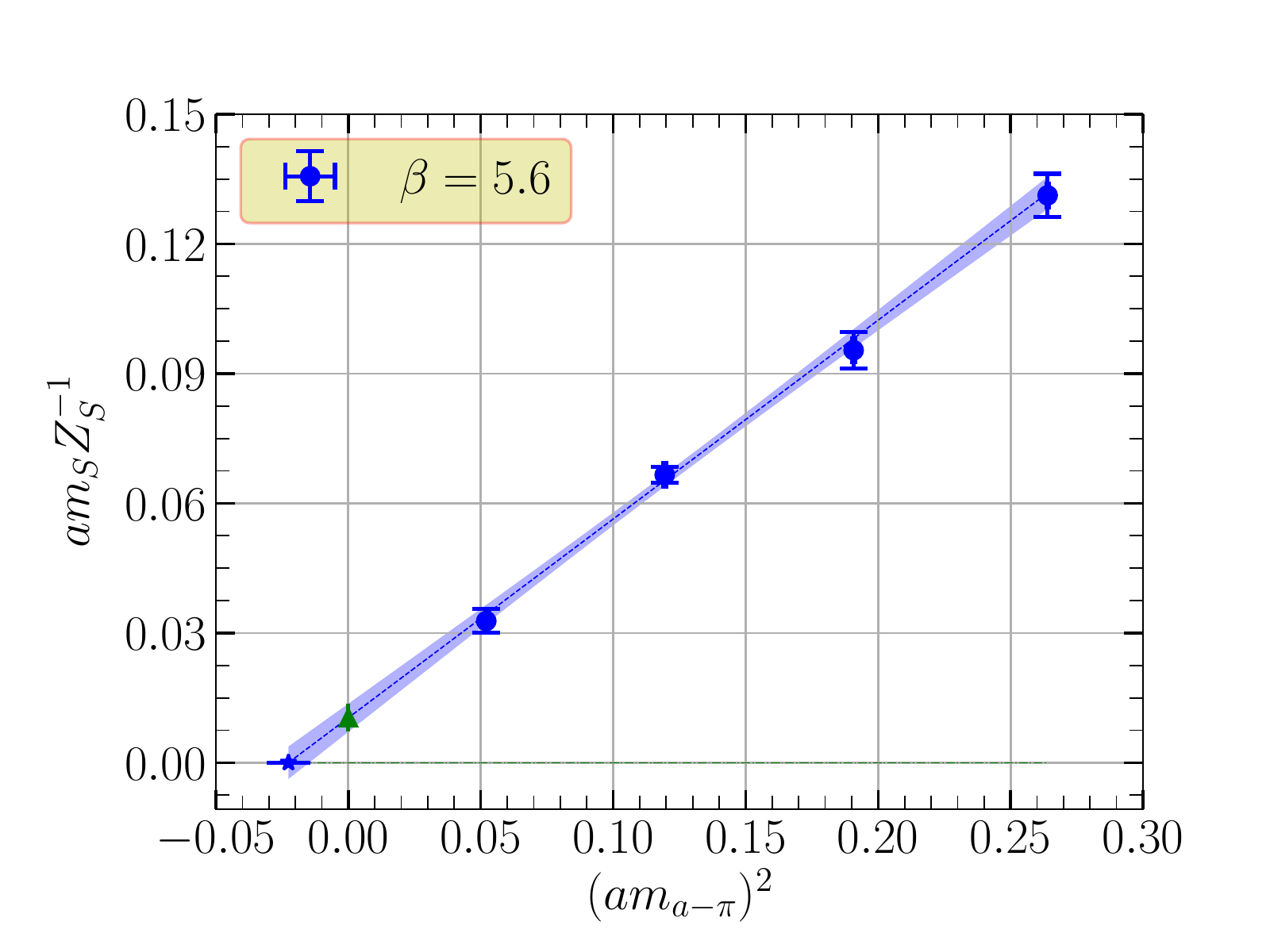}}
\caption{Chiral limit and determination of the remnant gluino mass at
$\beta=5.6$. All quantities are in lattice units.}
\label{Kcdm}
\end{figure}

The numerical results for the subtracted gluino mass from the Ward identities
and the adjoint pion mass squared in lattice units are shown 
for $\beta=5.6$ in Fig.~\ref{Kcdm}
together with their extrapolations towards the chiral limit.

In the continuum the subtracted gluino mass and the adjoint pion mass
should vanish at the same point. On the lattice, however, this is not the
case due to lattice artefacts. As an estimate for this discrepancy we
determine the value of the subtracted gluino mass at vanishing adjoint 
pion mass. This quantity is called the remnant gluino mass 
$\Delta(a m_S Z^{-1}_S)$, and it is expected to vanish in the continuum
limit. The values of the remnant gluino mass, obtained by taking an average
of the values calculated using the procedures explained above, are presented in
Tab.~\ref{RemnantGluino}.
\begin{table}[hbt!]
\begin{center}
\begin{tabular}{|c|c|c|c|c|c|}
\hline
$\beta$ & 5.4 & 5.45 & 5.5 & 5.6\\
\hline
$\Delta(am_SZ^{-1}_S)$ & 0.0334(48) & 0.019(12) & 0.0099(88) & 0.0103(33)\\
\hline
\end{tabular}
\caption{The values of the remnant gluino mass $\Delta(am_SZ^{-1}_S)$
obtained at four different values of the inverse gauge coupling.}
\label{RemnantGluino}
\end{center}
\end{table}
%

\subsection{Continuum limit}

\begin{figure}[hbt!]
\centering
\subfigure[The remnant gluino mass from separate extrapolations to the chiral
limit where $m^2_{\api}$ is zero, and the extrapolation to the continuum
limit.]
{\label{WIsCont2s}
\includegraphics[width=0.485\textwidth]{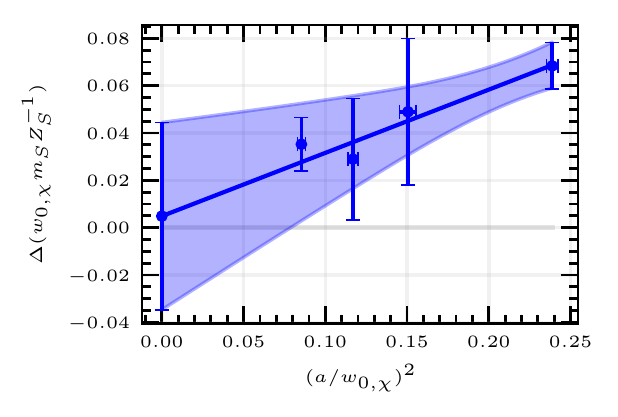}}\quad
\subfigure[The remnant gluino mass from a simultaneous chiral and continuum 
extrapolation. By construction, in this method the data points coincide with 
the error band.]
{\label{WIsCont1s}
\includegraphics[width=0.463\textwidth]{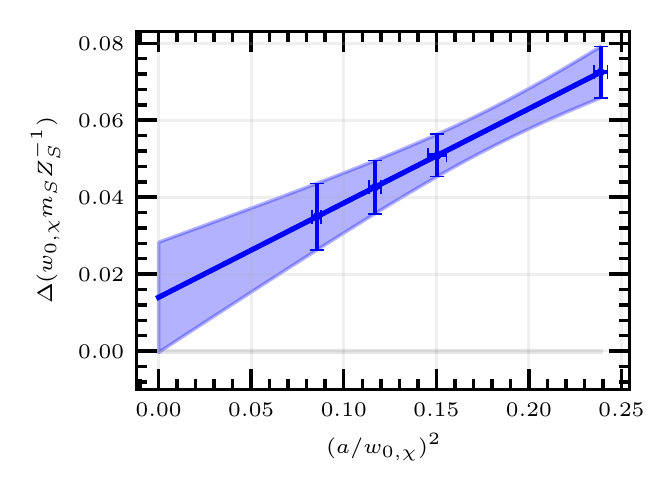}}
\caption{The remnant gluino mass $\Delta{(w_0 m_S Z^{-1}_S)}$ in physical
units $w_0$ as a function of the lattice spacing squared, and
its linear extrapolation towards the continuum limit.}
\label{WIsCont}
\end{figure}

The remnant gluino mass is a lattice artefact and should vanish in the
continuum limit $a \rightarrow 0$. It is therefore a quantity to check on whether
supersymmetry is recovered or not. Concerning the dependence of the
remnant gluino mass on the lattice spacing, arguments based on partially
quenched chiral perturbation theory suggest that the remnant gluino mass 
is of order $a^2$ at $m^2_{\api}=0$ \cite{Farchioni:2001wx}.
In order to investigate this relation, the remnant gluino mass has to be expressed in
physical units. Our choice for the scale is the Wilson flow parameter $w_0$, 
which is defined through the gradient flow \cite{Ali:2018dnd}. 
We use its values extrapolated to the chiral limit, $w_{0,\chi}$.
Similarly the lattice spacing is represented by $a/w_{0,\chi}$.
Our numerical results for the remnant gluino mass as a function of the
lattice spacing and its extrapolation towards the continuum limit are
shown in Fig.~\ref{WIsCont}.
The data points in Fig.~\ref{WIsCont2s} show the results from separate
chiral extrapolations for each lattice spacing and the corresponding
extrapolation to the continuum limit.
The extrapolation to the continuum and the error of this extrapolation
are obtained by means of parametric bootstrap with linear fits.
On the other hand, Fig.~\ref{WIsCont1s} is obtained by means of a
simultaneous fit of the dependence on the hopping parameter and the
lattice spacing~\cite{Gerber:2019thesis}.

The remnant gluino mass in the continuum limit
is compatible with zero within one standard-deviation, confirming the preliminary 
results present in 
Ref.~\cite{Ali:2018fbq} with only two data points. 
Lattice artefacts vanish in the continuum limit as expected, 
and supersymmetry is recovered in the chiral and continuum limits,
in agreement with our findings from the mass spectrum \cite{Ali:2019agk}.

\section{Conclusion}

In this paper we have presented numerical results of an analysis of SUSY
Ward identities in $\mathcal{N}=1$ supersymmetric Yang-Mills theory on the
lattice with gauge group SU(3). Contact terms and $O(a)$ lattice artefacts
in the Ward identities have been controlled by suitable choices of time-slice
distances. Ensembles of gauge configurations at four different values of the
lattice spacing and various hopping parameters have been analysed, 
allowing us for the first time to perform an extrapolation to the continuum 
limit, where the lattice artefacts vanish. The remnant gluino mass has been 
extrapolated in two alternative ways, on the one hand by extrapolating to the 
chiral limit at each lattice 
spacing separately and then to the continuum limit, and on the other hand by means
of a simultaneous extrapolation to the chiral and continuum limit.
With both extrapolations the lattice artefacts in the subtracted gluino mass appear
to scale to zero as of order $a^2$ in agreement with the theoretical expectations.
Our findings support the validity of SUSY Ward identities and the restoration of 
supersymmetry in the continuum limit. 

\section*{Acknowledgments}

The authors gratefully acknowledge the Gauss Centre for Supercomputing
e.\,V.\,\linebreak(www.gauss-centre.eu) for funding this project by providing
computing time on the GCS Supercomputer JUQUEEN and JURECA at J\"ulich
Supercomputing Centre (JSC) and SuperMUC at Leibniz Supercomputing Centre
(LRZ). Further computing time has been provided on the compute cluster PALMA
of the University of M\"unster. This work is supported by the Deutsche
Forschungsgemeinschaft (DFG) through the Research Training Group ``GRK 2149:
Strong and Weak Interactions - from Hadrons to Dark Matter''. G.~Bergner
acknowledges support from the Deutsche Forschungsgemeinschaft (DFG) Grant
No.~BE 5942/2-1. S.~Ali acknowledges financial support from the Deutsche
Akademische Austauschdienst (DAAD).


\end{document}